Review

# Capturing the geometry, function, and evolution of enzymes with 3D templates

Ioannis G. Riziotis[1*], Janet M. Thornton[1]

[1]European Bioinformatics Institute (EMBL-EBI), Wellcome Genome Campus, CB10 1SD, Cambridge, UK

[*]*To whom correspondence should be addressed:* riziotis@ebi.ac.uk, tel.: 0044 77 490 32067

Running title:

**3D templates for enzyme structure characterisation**

**Abstract**

Structural templates are 3D signatures representing protein functional sites, such as ligand binding cavities, metal coordination motifs or catalytic sites. Here we explore methods to generate template libraries and algorithms to query structures for conserved 3D motifs. Applications of templates are discussed, as well as some exemplar cases for examining evolutionary links in enzymes. We also introduce the concept of using more than one template per structure to represent flexible sites, as an approach to better understand catalysis through snapshots captured in enzyme structures. Functional annotation from structure is an important topic that has recently resurfaced due to the new more accurate methods of protein structure prediction. Therefore, we anticipate that template-based functional site detection will be a powerful tool in the task of characterising a vast number of new protein models.

**Statement of significance**

A review covering the use of 3D templates to identify conserved motifs in proteins, focusing on enzymes. We overview relevant literature and propose some new methodologies and implications, currently under development in our group. We anticipate that templates will aid in the functional annotation of predicted protein structures, since the advent of *AlphaFold DB*. We also discuss template-based methods to investigate the evolution of catalytic function, aiming to better understand catalysis and design new enzymes.

## Introduction

The mechanistic description of enzymatic catalysis requires firm understanding of the geometry of chemically important groups, whose relative spatial arrangement is well-defined at each step of the catalytic mechanism. Enzymes utilize a restricted ensemble of amino acids to perform all known catalytic reactions[1], with 3D conformation of the active site being critical for catalysis to occur. In a recent work we show that the geometry of active sites across a large set of diverse enzyme families is highly conserved, with conformational variation and flexibility being observed to various degrees amongst different enzymes[2].

Active sites are typical examples of functional sites that can be represented by 3D signatures, known as structural templates[3]. These usually consist of residue coordinates (a subset of main or side chain atoms, or average coordinates) and a set of geometrical or physical/chemical constraints to permit tunable matching specificity[4,5]. Special tailor-made algorithms are used to query protein structures and find template matches, working as a structural equivalent of sequence motif searching methods[6–9]. Templates have been used extensively for the identification of functional sites in protein structures[10]. More specifically, sites facilitating DNA binding[11], metal coordination motifs[12,13], catalytic sites[5,14] and ligand binding cavities[15,16] have been studied in our group, and template-matching methods, able to identify such sites have been incorporated in functional annotation pipelines[17]. During the era of structural genomics[18], protein functional annotation was a trending topic[19]. Fast-forwarding to the present day, the emergence of sophisticated protein structure prediction algorithms[20] is reviving the need for functional annotation from structural models, therefore revisiting templates for this purpose is important, though not a trivial process. Two challenges must be faced: how to generate template libraries, and how to query structures with templates, efficiently with speed and accuracy. In this review we focus on catalytic templates, aiming to address these challenges, though the principles can be transferred to templates representing all types of functional sites.

Furthermore, we present applications of templates in the context of enzyme evolution, discussing phenomena like functional convergence and divergence[21], catalytic site plasticity and flexibility. Understanding enzyme evolution is critical for designing enzymes that perform novel functions, and templates can aid to infer evolutionary relationships. We appreciate that enzyme engineering through modification of existing enzymes requires knowledge of how catalytic function has evolved in 3D, a topic beautifully discussed in the extensive body of work by Dan Tawfik[22,23]. In addition to reviewing the work done in structural templates so far, this paper presents some preliminary results from our ongoing work on enzyme evolution, using the well-studied catalytic triad of serine proteases as a case study[24] and as proof of concept for some proposed methodologies involving catalytic templates.

**Generating templates**

*Definitions of templates*

We define two types of templates: **coordinate** templates (model coordinates and a set of matching constraints), and **fuzzy** or "feature" templates. **Coordinate templates** consist of a set of atom coordinates, specifying a defined geometry, and each atom type is subject to constraints such as matching fuzziness (e.g. only side chain C atom, or any non C atom can be matched). These coordinates are used to generate distance constraints, which are applied when matching a target structure. The quality of the match is measured by the Root Mean Square Deviation (RMSD) of corresponding atomic positions[4,25]. Algorithms used for matching of direct templates will be discussed further below. **Fuzzy templates** on the other hand do not necessarily include hard-coded atom coordinates, but instead a set of geometrical and/or physicochemical parameters (e.g. inter-residue distances, charge, hydrophobicity etc.). The following paragraphs list a number of examples of templates of both types and the methods of their extraction, which can be either automated or semi-automated (manual/hybrid). Most studies involving template libraries implement an automated pipeline to extract motifs from protein structures, usually with additional sequence-derived information. This allows for large libraries to be generated quickly, covering a large functional space, but this can come with a cost in accuracy due to a variety of causes (e.g. no manual curation of templates) and usually, reviewing of the generated templates is mandatory.

*Template extraction methods*

Fetrow and Skolnick in their popular work[26], introduced the term "Fuzzy Functional Forms" (FFFs) to refer to feature (distances, angles) templates of relaxed constraints, generated manually for two enzyme families. These were used successfully to predict function by detecting catalytic centers both in high and low-resolution structures. This method was later automated by Arakaki et al., who developed "Automated Functional Templates" (AFTs), an upgraded version of FFFs. Their library comprised 593 indirect templates for 162 enzymes[27].

Enzyme function identification was also explored by Meng et al.[28], who show that consensus templates for two enzyme families (enolases and haloacid dehalogenases) coded in the SPASM format (see below in the Matching Algorithms section), consisting of a set of conserved residues, can successfully re-assign enzymes of a training dataset to the superfamily they belong. The same group, two years later, presented the automated version of this work (GASPS motifs)[29]; these motifs are culled from a set of conserved residues able to best identify the correct function in a training set. Residue conservation is also exploited in the works of Kristensen et al.[30] and Ward et al.[31] where catalytically critical residues are identified via Evolutionary Trace (ET); a library covering 98 enzyme families was generated by Kristensen et al., with these templates being ~80% successful to identify catalytic function in benchmarks. Catalytic site identification is also described by Nebel[32] who generated consensus 3D templates, built using rigid bound groups as a reference (porphyrin-bound proteins are explored as a use case). Another example of automatically extracted templates is presented by Liang et al.[33], whose method rely on gathering structural information from the environment of conserved sequence motifs; this information is used to construct templates in the form of physicochemical fingerprints. A template in this library is a hard-coded consensus of features and parameters instead of actual coordinates. Functionally important clusters of residues can also be identified by representing protein structures as graphs, an approach first devised by Artymiuk et al. in the ASSAM program[34] (described in the next section). DRESPAT by Wangikar et al.[35], also implements a graph-based method to break a protein down into small structural patterns (subgraphs) and then search a query structure for 3D motif matches in the form of common subgraphs. A similar concept is described by Laskowski et al., where a protein is broken into multiple direct templates of a few residues (reverse templates), and structures are queried with all of them, to identify function[36]. Functional characterisation has also been explored by Jambon et al. in the program SuMo[37], whose principle is based on comparing two structures and identifying common local 3D similarities.

Within our own group, Wallace et al., used the catalytic triad as a testing active site to generate functional atom templates that were able to distinguish catalytic from non-catalytic entities in protein structure databases[38]. This pipeline was further generalized a few years later by Torrance at al., who generated direct templates from 147 families in the Catalytic Site Atlas (CSA)[39,40], and validated their ability to identify members within the correct homologous family[5]. In that work, a representative template from each CSA family was extracted automatically using RMSD criteria. All CSA families include a representative structure where catalytic residues are manually curated, therefore this method of template extraction can be classified as hybrid, since it combines programmatic generation of templates and a knowledge-based definition of the active sites.

We have also considered binding sites; Jones et al. modelled helix-turn-helix motifs in 3D templates to identify DNA binding sites[11], while metal binding sites were explored by Torrance et al.[12] and Andreini et al.[41,42] in 2008. Similar work has been done by other research teams: Goyal and Mande[13], created metal-coordination templates by extracting metal-ligand neighbouring residue parameters such as inter-residue distances, $C_\alpha$ and $C_\beta$ plane angles and volume of the binding site. Furthermore, Zhao et al. generated fuzzy templates from non-homologous nucleotide-binding proteins. These are useful for identifying co-factor (NAD, FAD, ATP etc.) binding sites[25]. The structural features here are affinity potentials within the nucleotide binding cavities, which are used to generate a template that resembles a consensus pseudo-ligand (a physicochemical signature common among these cavities). Match-searching of the template ligand is performed using molecular docking software. Lastly, Chang et al. introduced Protemot[43] in 2006, a web server for ligand binding site prediction. This is again based on coordinate templates, generated automatically by the extracting residues in close proximity to ligands bound in crystal structures from the PDB[44]. This method initially produced a high rate of false positive results, but was later refined by introducing a sequence-order constraint to the templates[45], that significantly reduced this rate.

*3D variation-aware templates*

The limitation of the template library generated by Torrance et al. lies in the fact that a single template is generated for an enzyme family, averaging out any structural variation within homologous active sites[46]. This applies to the striking majority of libraries involving some form of consensus templates: Fuzziness caused by structural variation is only captured by setting relaxed matching criteria in the matching algorithms, leading to a high rate of false positive results, and reducing accuracy. What if different conformational states of an active site could be captured by multiple templates? Fig. 1 presents this concept, exploiting GTPases –a use case thoroughly examined in our previous work (See ref. 2). The methodology is simple: catalytic residues for all homologous enzymes of a (Mechanism and Catalytic Site Atlas (M-CSA)) family are extracted and iteratively superimposed[47] over their functional atoms[5]. Structures are then clustered by constructing a hierarchical dendrogram using pairwise RMSD as a metric, and the tree is pruned to derive structural clusters (adjusting pruning height will lead to a coarse or fine clustering). In GTPases (Fig. 1), three major clusters can be distinguished, each of which corresponds to the binding of a different analogue to the native ligand as this is transformed during catalysis (Cluster 1: Transition state analogue, GDP.AlF$_3$; Cluster 2: Substrate analogue, GSP; Cluster 3: Product, GDP). For each cluster we generate a representative template, derived from the site with the closest 3D similarity to the average cluster coordinates. The result is three templates, each one corresponding to a major conformational state of the active site during the progression of catalysis. We see that the primary differences in geometry lie in the Arg residue which directly interacts with the *β* and *γ* phosphate groups of the ligand by forming an H-bond with an oxygen atom between them. Other residues such as Gln/Asn and bottom right Asp/Glu also exhibit some variance, which is captured by the templates. Important to note here also is the fuzzy residue matching that is allowed in the templates (Asp-Glu, Asn-Gln, Ser-Tyr-Thr can be matched interchangeably) and fuzzy atom matching (e.g. endpoint Asn/Gln N and O atoms can be matched interchangeably[48]). This approach, which we are currently developing, will allow for a much more comprehensive and robust catalytic template library, that can also assist in understanding catalysis and in the design of novel enzymes.

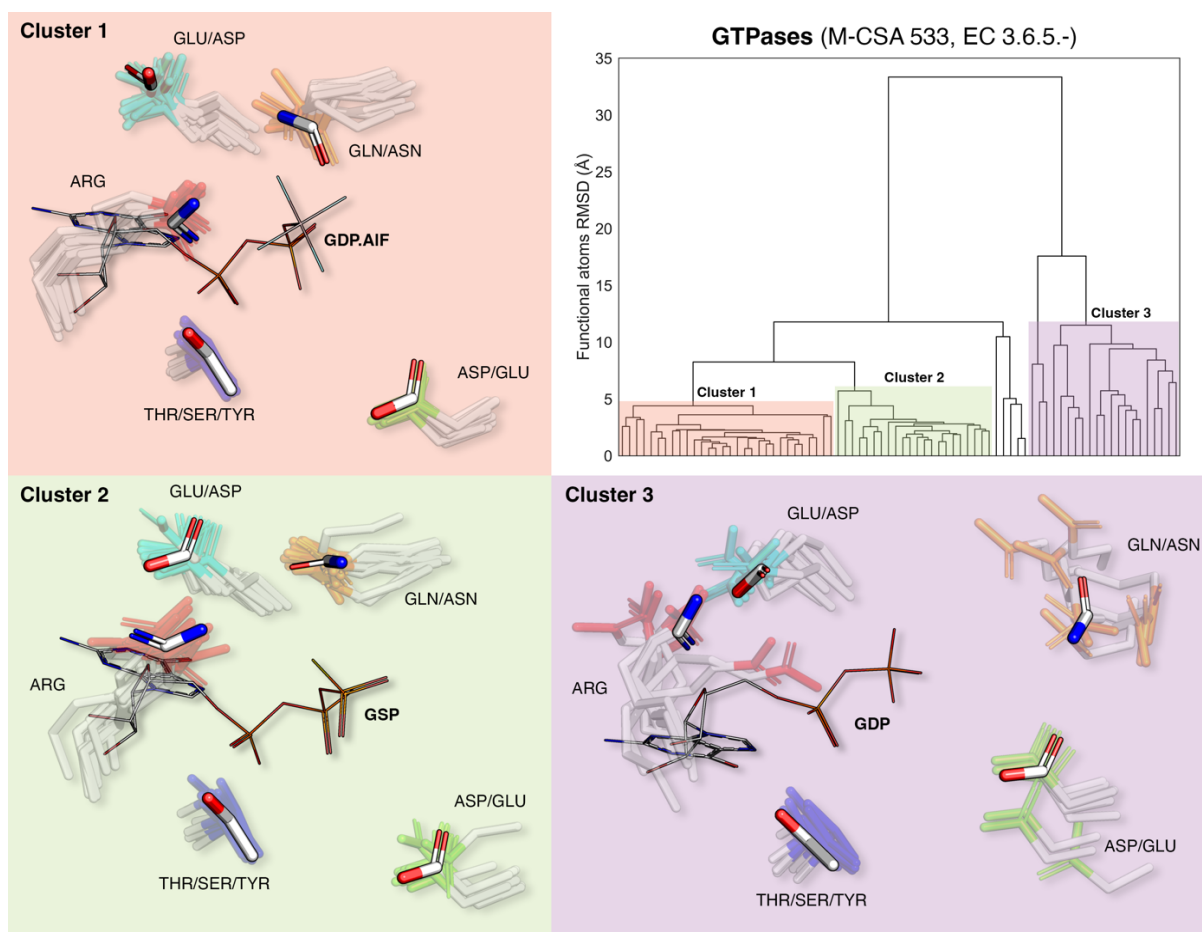

*Fig. 1: Multiple 3D templates to cover structural variation in the active sites of GTPases. Catalytic residues from enzymes within the homologous family are superimposed over their functional atoms, and clustered by constructing a hierarchical dendrogram (upper right panel). Major clusters (presented over distinctly coloured backgrounds) are derived by dynamically pruning the dendrogram. Side chains of catalytic residues are shown as transparent sticks, with their functional atoms coloured differently according to the group of aligned residues they belong. The 3D template for each cluster is a representative active site calculated as the cluster member of closest 3D similarity to the average coordinates of the cluster, and is shown as opaque sticks, coloured by atom type (white: C, red: O, blue: N). Alternative residues that can be matched by template atoms are labeled accordingly. For each cluster, a representative bound ligand (Cluster 1: GDP.AlF3 – transition state analogue; Cluster 2: GSP – GTP substrate analogue; Cluster 3: GDP – product) is shown as light sticks, coloured by atom type. 3D models were prepared in PyMol[49].*

## Matching algorithms

In this section we discuss a number of template matching algorithms. This is merely a subset of all algorithms in literature, but they comprise a sample representing important underlying principles for template matching. One of the first principles utilized for 3D motif matching was search via subgraph isomorphism, implemented in the program ASSAM[34] by Artymiuk et al. In this algorithm side chains are abstract and are represented by pseudoatoms (graph nodes), with their relative position represented by distances (graph edges). Ullman's subgraph isomorphism is then used to find occurrences of the template (as subgraphs) in a protein structure graph. Stark and Russell also used inter-residue distances to find template matches in PINTS[19], also introducing an atom-count independent confidence score, to be used complementarily with RMSD for assessing the odds of finding a template match by chance. Similar principles (depth-first search using Euclidean distances) are implemented in the SPASM-RIGOR[3] software suite: here, residues in the template are modelled by a $C_\alpha$ atom and a pseudo-atom representing the side chain centre of mass. Fuzzy residue and atom matching are supported in this method, and the output is all substructures satisfying the template constraints, with the drawback being the lack of a statistical significance metric for each match. The working principle is identical between SPASM and RIGOR, with the former being used for searching a user-defined template in a set of query structures, and the latter for querying a single structure with a library of pre-defined templates. Higher flexibility in template nomenclature is allowed in the program TESS[25], which uses geometric hashing for template searching, an algorithm used as of then in computer vision. Templates of similar (and also more flexible and robust) nomenclature are also used by Jess[4], the successor of TESS; this is based on a k-d tree method to represent a structure and iteratively search for matches of the constraints coded in a template. This provides order-of-magnitude increase in speed compared to TESS since it does not require severe preprocessing and storing of the query structure. Also, Jess uses a semi-empirical significance score for the outputted template-match superposition, calculated through model RMSD distributions of real structural data. Jess and its templates will be used in the examples below. Lastly, one of the latest motif matching algorithms reported by Brittich et al. applies an inverted-index search principle to prioritize structural database searching, by breaking templates in two-residue combinations, significantly improving efficiency and speed[50].

Distinguishing true from false positive matches is a significant problem in template matching. This heavily relies on template specificity, which further relies on the defined constraints and number of residues/atoms (lower number or residues will lead to more general templates, thus more potential hits). For instance, in enolases, five-residue templates used by Meng et al.[28] seem to be robust enough to correctly assign these enzymes to their superfamily, but for other superfamilies, the optimal residue number might be different. The question underlying this is the following: What is the optimal specificity, so that we don't reject matches of biological significance, for example those that result from convergent evolution? Furthermore, structural variation within the active site is difficult to capture when templates are over-specific; a tradeoff is to loosen matching criteria (e.g. a relatively high pairwise matching distance threshold or a high RMSD threshold). However, this comes at a cost in computing speed, with higher distance and RMSD thresholds resulting in slow algorithm runs[4], and again a higher false/true positive ratio. A feasible solution to these problems can potentially be provided by the recently introduced parametric templates[51], where individual atoms can be weighted according to arbitrary criteria.

As discussed further above, there are confidence metrics to assess meaningfulness of matches. Stark et al. describe an *a priori* method for measuring statistical significance of calculated RMSD of fitted motifs in query structures, in a fold-independent manner[52], while Jess implements a semi-empirical metric. It is evident that this is a significant challenge. For example, Protemot[43] suffered reliability issues caused by a high false positive rate. However, its back-end algorithm was optimised by applying a sequence-order constraint[45] that demanded matching residues to be in the same sequential order as in the template. Although this might provide robustness in identifying matches in proteins linked by evolution, it is severely limited in detecting convergent active site geometries or functional motifs where residues might have been substituted by others in different position in sequence.

**Applications**

Functional annotation has always attracted great interest in protein science, since experimental functional characterisations are time-consuming and rarely unbiased and therefore several computational approaches have been developed for this[53–56]. Functional identification pipelines using templates have been released, including but not limited to ProFunc[36], COFACTOR[57], SuMo[37] etc. These found huge application during the rise of structural genomics two decades ago, which added more than 15,000 experimentally determined structures to the PDB, plenty of them being functionally uncharacterised[58]. It is easily predictable that similar work will re-emerge, with the latest breakthrough in protein structure prediction. The public release of the AlphaFold database[20,59] has delivered millions of modelled structures, the majority of which are of unknown function. Therefore, the importance of functional annotation is rising again, and templates are a great tool to facilitate this.

Templates also find application in studying the evolution of proteins. Inference of evolutionary relationships is a non-trivial task in cases where sequence and structure have diverged to the point where it is impossible to distinguish random (non-functional) similarities from convergent evolution events and from similarities due to the presence of a common ancestor[21,26,60]. In such cases, templates identifying local structural similarities can provide some insight into evolutionary relationships. We are particularly interested in convergent and divergent evolution in enzymes, as well as structural substitution events in active sites (plasticity) that results in rescue of catalytic function when major mutation events occur during evolution. A detailed review of the latter phenomena is described and illustrated in a survey by Todd et al.[61]. In the following paragraphs we use the well-studied Ser-His-Asp catalytic triad of serine proteinases to describe some applications of templates in exploring the evolution of enzyme catalysis.

**Functional convergence and divergence**

It has been shown that convergence in the function of evolutionary unrelated enzymes, particularly among large enzyme families, occurs frequently[60,62]. Similarly, due to the restricted number of distinct protein folds in nature, enzymes catalysing slightly[63] or even radically different reactions might adopt a similar overall structure, with differences only located within the catalytic centre. This is often the outcome of divergent evolution[64], usually after gene duplication and subsequent selection in the gene copies.

Templates are powerful tools to detect and investigate such phenomena. In functional convergence, similar relative spatial arrangement of functional groups can be detected in active sites. Subtle 3D differences in functional divergence on the other hand, will correspond to a partial match of a template to the active site (e.g. matching of 3 out of 5 residues), with non-matching functional atoms reflecting the residues that have changed during time and resulted to functional shift (change in mechanism[21] or change in substrate specificity[65,66]).

Serine proteases are a well-known example of functional convergence in enzymatic function. Russell, using the catalytic triad as a use case along with some ligand binding sites, described side chain conformational patterns observed frequently as a consequence of convergent evolution[67]. These patterns are essentially 3D templates that can identify conserved geometries of functional atoms. Querying a protein with two templates of the same active site, one consisting of backbone and one of functional atoms, can constitute a simple method to identify cases of functional convergence. The principle is the following: Functional atoms matches in a structure, that don't have a corresponding backbone atoms match, will indicate a conserved geometry in the side chain, while the residues contributing these side chains might come from completely different sequence positions.

Using the catalytic triad as an exemplar, we present some indicative results of a rudimentary pipeline like this, in Fig. 2. Here, a non-redundant set of PDB structures (95% max sequence identity) is queried using consensus Jess templates of the Ser/His/Asp catalytic triad extracted from carboxypeptidases (M-CSA entry 5, EC 3.4.16.6). The dataset was first queried with a backbone template, and subsequently with a functional atoms template (same software parameters used in both searches). The functional atoms template allows fuzzy matching of Ser with Thr and Tyr, as well as Asp with Glu; also, to minimise the number of false-positive hits, inter-residue distances in the matches must not differ from the corresponding template distances by more than 1Å. From the putative hits (functional atoms matches not having a corresponding match in backbone), we filtered the structures annotated with hydrolase activity (EC 3.-.-.-); 3 hits were cherry-picked by hand and are presented in individual panels in Fig. 2a. It is clearly seen that catalytic triad geometry is observed within different structural contexts, as seen from the secondary structure variety among the hits. Superposition of the hits over their functional atoms (bottom right panel), reveals that these are actually well conserved in 3D, whilst their respective $C_\alpha$ atoms (orange spheres) are scattered in various directions. This shows the many configurations and possibilities to form a catalytic triad functional motif, by residues from completely different positions, both in sequence and in the fold. These results demonstrate the power of templates to identify convergence; further evidence that these proteins are not evolutionarily related is provided by the CATH[68] domain annotations of each triad site, as well as by the multiple sequence alignment of the sub-sequences containing the triad residues (Fig. 2b), which shows no evident homology.

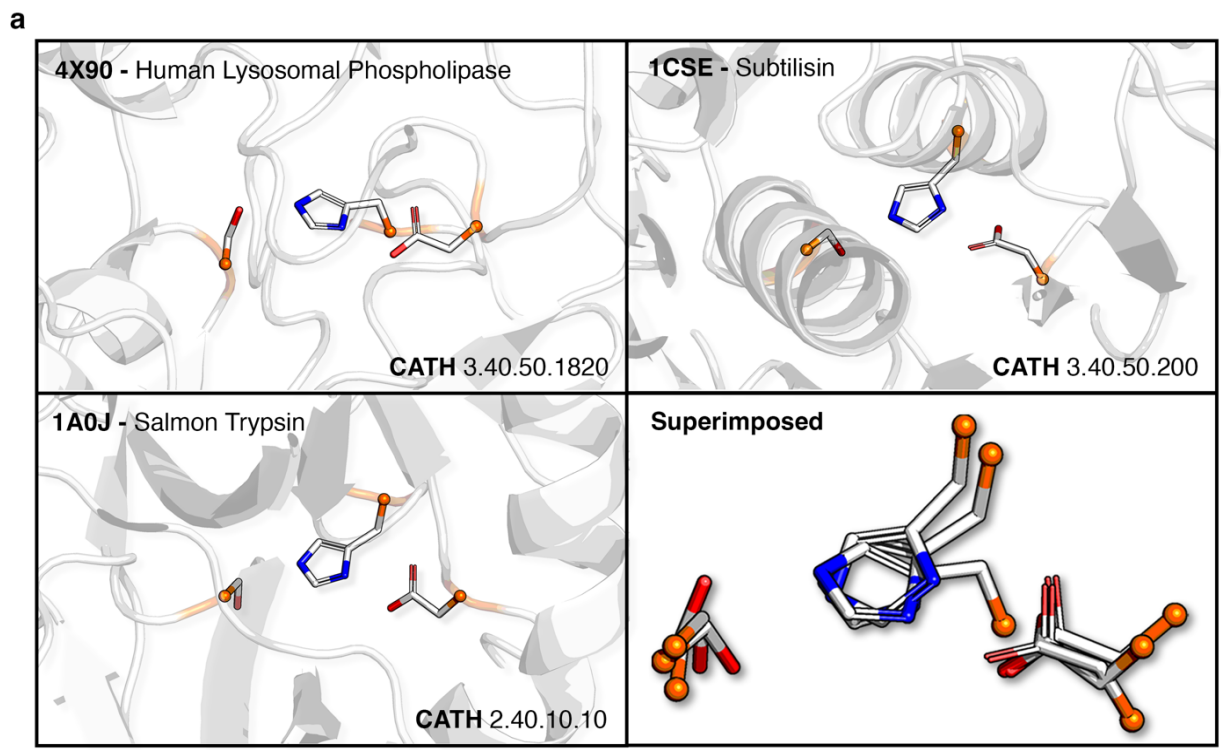

*Fig. 2: Examples of convergent catalytic triad 3D motifs, identified by a template consisting of functional atoms. a: Top panels and bottom right panel show each example, with side chain atoms coloured according to atom type and main chain coloured orange. A part of the parent structure in each example is drawn as cartoon to indicate the different structural context of these motifs. Bottom right panel shows all hits superimposed over the template in a stick representation, with side chains coloured by atom type (white: C, red: O, blue: N) and $C_\alpha$ atoms shown as orange spheres to indicate the geometrical heterogeneity of the backbone and geometrical/chemical conservation of the functional groups. 3D models were prepared in PyMol[49]. b: Multiple sequence alignment of sub-sequences containing the catalytic triad residues (shaded in pink), in the three example proteins. Matched residues in the alignment are shaded in purple, as illustrated in JalView[69].*

## Active site plasticity

Catalytic plasticity refers to the conservation of structure and function, especially within the active site, after numerous amino acid changes during evolution. Catalytic sites are under evolutionary pressure, therefore the effects of functionally deleterious mutations might be reversed by the deployment of other residues. The result is two active sites from evolutionarily related enzymes, with similar functional atoms' adopting similar geometries, but whose residues do not align perfectly in sequence. Todd et al. present a collection of cases culled from the literature[61], where various types of plasticity are exemplified. Rescue of function during evolution can be the outcome of –among others– spatial or functional substitution; the former refers to the use of a similar residue from a different position in sequence while the latter is the contribution of the same functional atoms by a residue of different chemical structure (e.g. O atom contributed by a Thr in one case and a Tyr in another case). Of course, there are other phenomena that lead to re-establishment of function, such as circular permutations and divergence followed by convergence, but here we focus on the first two mentioned.

We are again using the catalytic triad and present two simple case studies, identified by querying a non-redundant PDB set (95% maximum sequence identity)[70] with the [Ser/Thr]-[His]-[Asp/Glu] functional atoms template. Fig. 3 illustrates the first case, which a catalytic triad, detected in two bacterial carboxylesterases (EC 3.1.1.-). Both enzymes perform similar functions, fold into a common α/β architecture and belong to the same CATH superfamily, therefore are homologues. The template search identified a catalytic triad configuration in these structures; both are consistent with the literature and are responsible for hydrolysing a carboxylic ester into an alcohol and a carboxylate moiety[71,72]. The catalytic triad of the *E. coli* homologue (shown in pink in the figure) is the standard Ser-His-Asp triad found in most serine proteases, however, in the *B. subtilis* homologue, the Asp residue has been substituted by a Glu in the same sequence position. This is a case of functional substitution, where the function of the enzyme is rescued despite the mutation, and the geometry of the functional groups is almost perfectly retained. Asp and Glu share the same carbonyl functional group, and this is captured by the template, since we allow Asp-Glu to be matched interchangeably.

The second noteworthy case is again a pair of proteases, a human and a bacterial one, shown in Fig. 4. This case underscores the question: can we infer evolutionary relationships between proteins with the aid of structural templates? These two examples have a matched Ser-His-Asp catalytic triad each, both known to be functional[73,74], in the centre of the molecules (Fig. 4a). The triads have almost identical conformation, and as seen in superposition (Fig. 4b), the secondary structure around each share some similarity, although the outer shells of the structures are completely different (as seen in their CATH codes). This particular case was identified by template matching and subsequent multiple sequence alignment of the local sequences containing the matches motifs (Fig. 4c). The pair only has the Ser residues of the triads aligned in sequence, which are located on a C-terminal conserved motif in both proteins. The His and Asp residues on the other hand, come from completely different sequence positions, and they are even in different sequence order, as clearly seen in Fig. 4c. Corresponding aligned residues with these two are shown in in yellow in Fig. 4a and Fig. 4c, and they are located in positions in the structure that are unrelated with function[75]. What these observations suggest, is that this is most probably a case of convergent evolution. Another scenario might be that they could share a distant common ancestor (that would either be before eucaryote/procaryote differentiation or a scenario of symbiotic bacteria). Having said that, if we assume an ancestral link here, the Asp or His residues have been substituted during evolution by residues in a different position in the sequence. Therefore, this might be a case of single or double spatial substitution.

Commonalities like the ones discussed in this section could only be found by a purely 3D approach like the use of templates, although more information is required to exactly infer the evolutionary history. Sequence alignment is proven to perform poorly in cases where sequences have diverged to the point where only some extremely conserved motifs would be aligned. However, template matching can find an implication in these cases, as it is sequence independent and can recognize motifs only conserved in 3D, even if these derive from residues with dissimilar chemical structure (in the case of functional substitution).

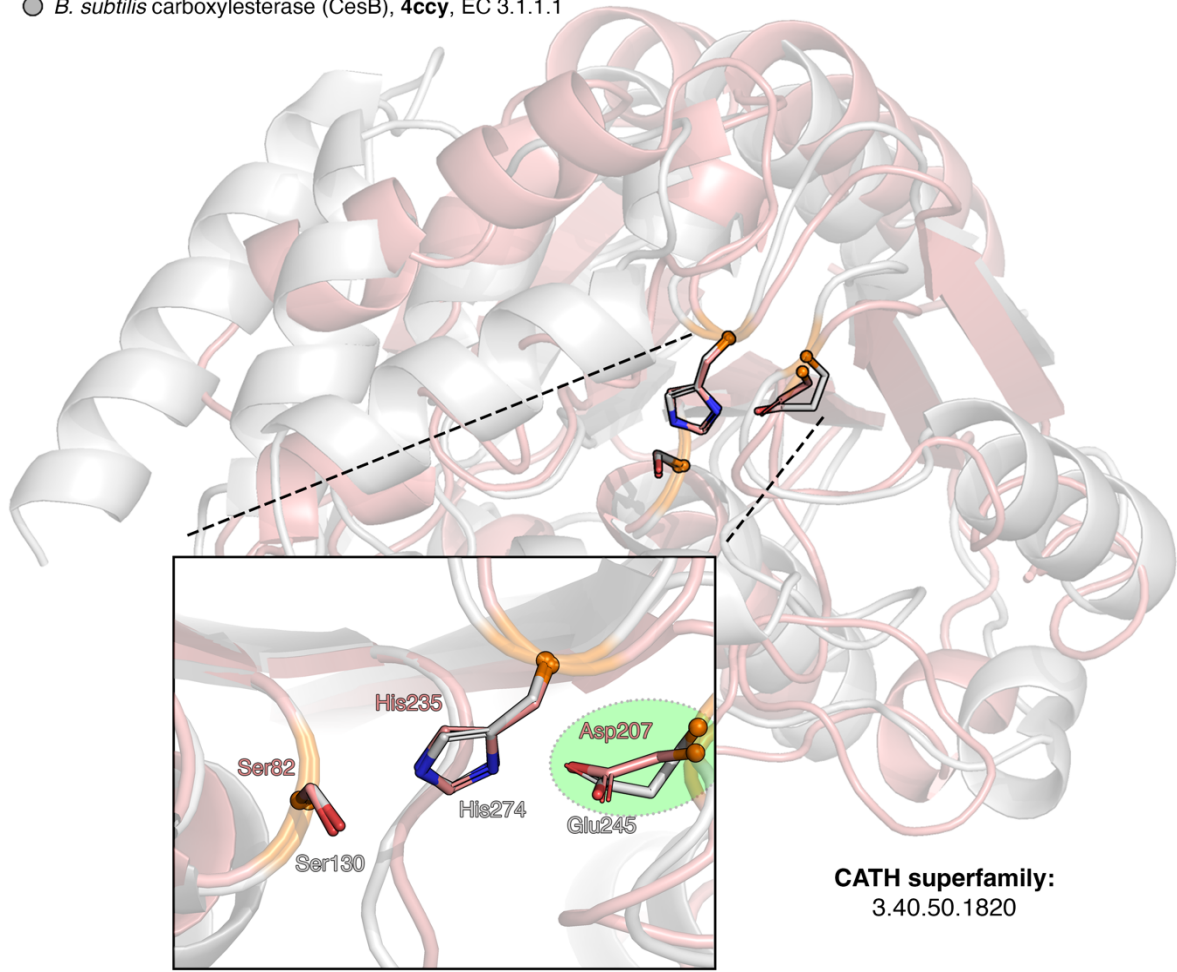

*Fig. 3: Functional substitution event in serine proteases captured by a 3D template. Two homologous bacterial carboxylesterases (EC 3.1.1.-) (PDB IDs: 1m33 (pink) and 4ccy (white)) are superimposed over the functional atoms of the template-matched catalytic triads. Inline panel shows the superimposed triads in a zoomed stick representation, coloured by atom type (C: white/pink, O: red, N: blue). Glu-Asp functional substitution is indicated by a green circle. 3D models were prepared in PyMol[49].*

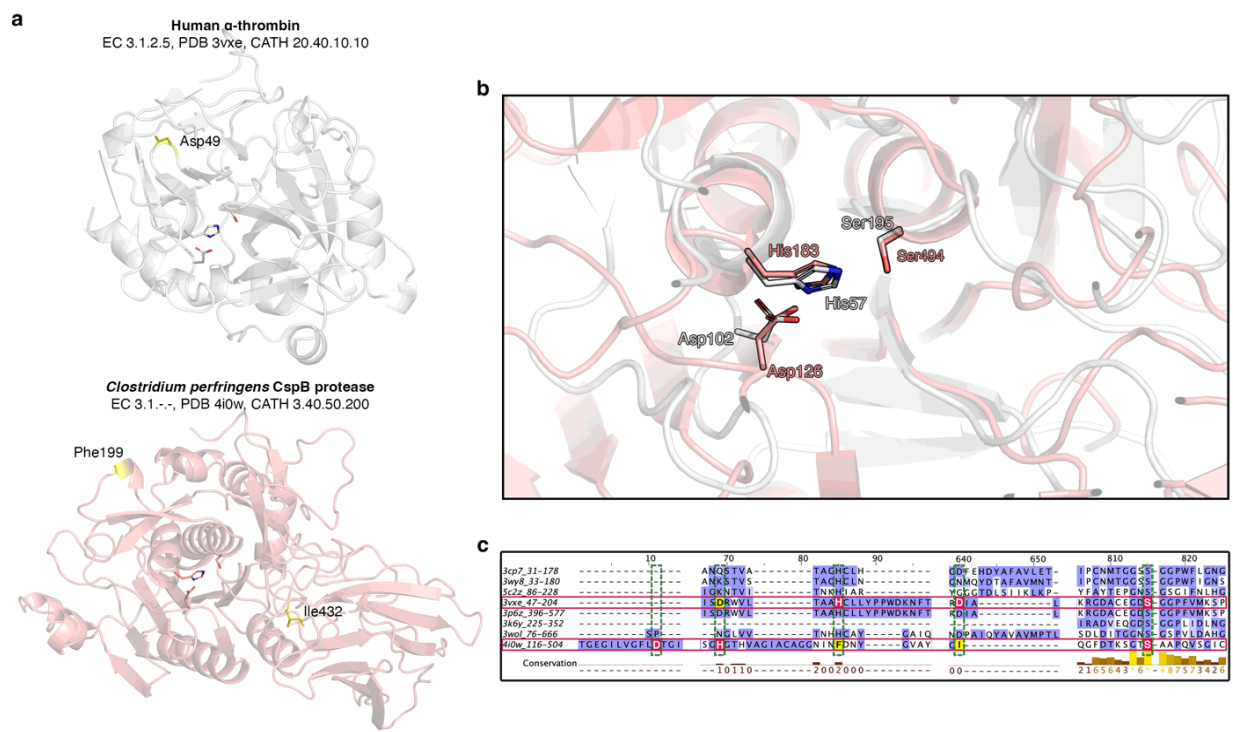

*Fig. 4: Catalytic triads captured by a 3D template, in two enzymes of similar function. a: A human (upper model, white, PDB ID 3vxe) and a bacterial (lower model, pink, PDB ID 4i0w) serine protease shown in cartoon representation, with the catalytic triad shown as sticks. Residues of the triad that align in sequence but not in structure (see panel c) are coloured yellow and labelled. CATH codes refer to the domains where the triads belong to. b: Magnification and superposition of the human and bacterial catalytic triads shown as sticks coloured by atom type. (C: white/pink, O: red, N: blue). Structural context is presented as cartoon to indicate a relative similarity in the fold within the substrate binding cavity. c: Multiple sequence alignment of a sample of serine proteases in which a catalytic triad motif was matched by the template with aligned positions shaded purple. The two proteases under investigation are marked by red rectangles. Residues of the catalytic triad are shown in red background and alignment columns (green dashed rectangles) indicate the order mismatch in two of the three catalytic residues (His and Asp). The corresponding aligned residues of the two misaligned residues are shown in yellow background and labelled in the 3D models of panel a. Note that only the subsequence blocks surrounding the catalytic residues are shown in the figure. 3D models were prepared in PyMol[49] and alignment was visualized in Jalview[69].*

## Conclusions – Discussion

This paper has reviewed template extraction pipelines and matching algorithms. Templates can be defined in different ways and derived using various approaches, some relying on automatic prediction of functionally important residues, and others being knowledge-based. Tailor-made matching algorithms have been developed for each type of template and almost all are limited by the same challenge: distinguishing true from false positives. Statistical analysis is usually performed to filter out matches occurring by chance[4,52], as well as adjustment of template specificity by adding/removing matching constraints. Here it is important to stress that the latter approach should be taken with care, as it entails the risk of creating over-specific templates (e.g. by including more residues/atoms in a coordinate template) that perform poorly in matching, especially when structural variation is involved.

Identification of functional centres in macromolecular structures is the primary goal of templates, as seen in several works presented in previous sections. This is an ongoing need, since the advent of –now millions[59]– of accurately (at least a significant fraction of them[76]) predicted 3D protein models, plenty of which lack functional annotation. Template-based annotation of these models can lead the way to further characterisation, this time experimental. Furthermore, this works as a positive feedback loop: new structures are functionally characterised (computationally or/and experimentally) and their functional sites are used to further enhance existing templates (or create new ones), making the libraries progressively more robust.

We also consider the use of templates to explore the evolution of enzymes. Phenomena like convergent, divergent evolution, functional promiscuity, moonlighting and functional site plasticity can indeed be identified and characterised through 3D templates. However, such procedures are rarely trivial; from the cases we presented in the previous sections, we saw that analysis of protein structures can help to elucidate the pathway of evolution, where sequences have diverged to such an extent that relationships are difficult to identify[77]. Template libraries can assist in inferring evolutionary pathways to new function, which is a critical piece in protein/enzyme design.

Lastly, we have presented some future ideas currently under investigation within our group. We introduced here the concept of multiple templates to cover conformational variation in flexible active sites. The extraction process is fully-automated and we plan to make the library available in the near future, both as standalone and as part of the back-end algorithm of ProFunc[17]. Furthermore, as seen in Fig. 1 and in our recent publication[2], large active sites like this can be broken down into self-contained entities or modules (for instance, a metal binding and a bond cleaving module, or a catalytic triad and an oxyanion hole). Our templates essentially represent the functional groups of catalytic residues, and by combining them with mechanistic information from M-CSA, we can define functional group combinations responsible for each mechanistic step. Such modules can serve as puzzle pieces to aid in modifying or designing catalytic centres in artificial enzyme design. Moreover, template-based analyses of evolutionary phenomena in enzymes are currently under systematic investigation; we anticipate these template methods to constitute a powerful set of tools to design, repurpose and modify enzymes, by mimicking how nature modifies enzyme active sites to retain function.

**Conflict of interest**

Authors declare no conflict of interest

**Acknowledgements**

The work was supported by: the EMBL International PhD Programme (IGR) and the European Molecular Biology Laboratory (JMT)

**Author contributions**

**IGR**: Conceptualisation, Methodology, Validation, Investigation, Visualisation, Writing – Original Draft preparation, Writing – Review & Editing

**JMT**: Conceptualisation, Supervision, Resources, Funding Acquisition, Project Administration, Writing – Review & Editing